\documentclass[epj,final]{svjour}

\usepackage{amssymb,amsmath,epsfig,graphicx}
\newcommand{\vmedia}{\langle \dot x_{\text{cm}}\rangle}
\newcommand{\vone}{\langle \dot x\rangle}
\newcommand{\Fext}{F_{\text{ext}}}
\newcommand{\Pout}{P}
\newcommand{\Fstop}{F_{\text{stop}}}
\newcommand{\Pmax}{P_{\text{max}}}
\newcommand{\sgn}{\text{sgn}}

\begin{document}
\title{Information and maximum power in a feedback controlled Brownian
  ratchet}
\author{{M. Feito\inst{1}
\thanks{feito@fis.ucm.es}
 \and F. J. Cao\inst{1,2 }
\thanks{francao@fis.ucm.es}%
}                     
%
%
}
\institute{
 Departamento de F\'isica At\'omica, Molecular y Nuclear,
Universidad Complutense de Madrid, Avenida Complutense s/n, 28040 Madrid,
Spain
\and
LERMA, Observatoire de Paris and CNRS UMR 8112, 61, Avenue de
l'Observatoire, 75014 Paris, France}
\date{Received: date / Revised version: date}
%
\abstract{ Closed-loop or feedback controlled ratchets are
Brownian motors that operate using information about the state of
the system.  For these ratchets, we compute the power output and
we investigate its relation with the information used in the
feedback control.  We get analytical expressions for one-particle
and few-particle flashing ratchets, and we find that the maximum
power output has an upper bound proportional to the information.
In addition, we show that the increase of the power output that
results from changing the optimal open-loop ratchet to a
closed-loop ratchet also has an upper bound that is linear in the
information. }
\maketitle

\section{Introduction}
Brownian ratchets have been studied in different contexts due to
their theoretical importance in non-equilibrium statistical
mechanics and their potential relevance for applications in
disciplines like nanotechnology, condensed matter or
biology~\cite{rei02,lin02,bie07,ser07}. Many studies deal with the
performance of these devices (see refs.~\cite{rei02,par02} for
comprehensive reviews) concentrating on open-loop ratchets, as
those obtained fluctuating an uniform external force (rocking
ratchets~\cite{ast94,mag93}), or an external asymmetric potential
(flashing ratchets~\cite{ast94,ajd93}), either randomly or
periodically. On the other hand, closed-loop or feedback
controlled  ratchets, as the so-called instant maximization
protocol~\cite{cao04} and the threshold protocol~\cite{din05}, use
information of the state of the system to operate. The feedback
ratchet of~\cite{cao04} has been recently proposed as an effective
model to describe the stepping motion of the two-headed
kinesin~\cite{bie07}. Other `information-dependent' rectification
mechanism have been recently proposed to model certain chemical
and biological systems~\cite{ser07}.

The previous works~\cite{cao04,din05,cao07} about closed-loop
ratchets focussed on the study of the flux and its maximization.
In particular, it has been shown that the increase of the flux
performance when the optimal open-loop control is changed to a
closed-loop control has an upper bound proportional to the square
root of the information used by the controller~\cite{cao07}.
\par

In this paper, we consider another measure of the performance, viz. the
power output, with the aim of getting further insight in the relation between
information and the increase of performance in a system with thermal
fluctuations. We oppose to the flux a constant load
force~\cite{sek97} in order to compute the potential energy gain by the
particles thanks to the action of the controller.
The generalization of the methods developed in~\cite{cao07} allow us
to obtain the relations between the maximum power output and the information
that the controller uses.

\section{The model}

The collective feedback ratchet that we investigate has two basic
ingredients, namely, $N$ Brownian particles and a controller.
The controller acts on the particles
switching on and off a potential $V(x)$ according to the control policy
and to the information received about the state of Brownian particles
through a noisy channel.
\par

Specifically, we consider $N$
overdamped
Brownian particles at temperature $T$ in a piecewise linear saw-tooth
potential
\begin{equation}
V(x)=
\begin{cases}
\frac{xV_0}{aL} & \mbox{if $0\leq \frac{x}{L}\leq a$}, \\
V_0-\frac{V_0}{1-a}\left( \frac{x}{L}-a\right) & \mbox{if $a< \frac{x}{L}\leq
  1$},
\end{cases}
\label{potential}
\end{equation}
of height $V_0$, asymmetry parameter $a$, and period $L$, \emph{i.e.}
$V(x)=V(x+L)$. The potential is switched on and off according to the
instant maximization of the center-of-mass velocity protocol (see
ref.~\cite{cao04}), which switches on the potential only when the net force
due to the potential on the particles would be positive.
In order to obtain work from the system operation we
oppose to the flow of particles an homogeneous static force $\Fext$; thus, the
total force acting on the particles when the potential is
on is $F_{\text{tot}}(x) = F(x) - \Fext$, with $F(x)=-V'(x)$, and
$F_{\text{tot}}(x) =  - \Fext$ when the potential is off. The state of
the system is described by the positions $x_i(t)$ of the particles that
satisfy the Langevin equations
\begin{equation}\label{langevin}
\gamma \dot x_i(t)=\alpha(t)F(x_i(t))-\Fext +\xi_i(t);\quad i=1,\dots,N,
\end{equation}
where $\gamma$ is the friction coefficient (related to the diffusion
coefficient $D$ through Einstein's relation $D=k_BT/\gamma$) and
$\xi_i(t)$ are Gaussian white noises of zero mean and variance $\langle
\xi_i(t)\xi_j(t^\prime)\rangle =2\gamma k_B
T\delta_{ij}\delta(t-t^\prime)$. The dichotomous function $\alpha(t)$
[$\alpha=0$ (potential off) or $\alpha=1$ (potential on)] implements the
action of the controller. The control policy uses
the information received from the system through a noisy channel that we
model with a binary symmetric channel~\cite{cov91}.
This channel passes the sign of the net force
\begin{equation}
f(t)=\frac{1}{N}\sum_{i=1}^N F(x_i(t))
\end{equation}
to the controller with an error probability $p$ known as the noise level of
the channel, so when $f(t)>0$ ($<0$) the controller switches on
(off) the potential with probability $1-p$. Therefore, the feedback protocol
and the noisy channel lead to the effective control policy
\begin{equation} \label{alpha}
\alpha_{\text{eff}}(t)=(1-p)\Theta(f)+p\Theta(-f),
\end{equation}
with $\Theta$ the Heaviside function [$\Theta (x)=1$ if $x>0$, else $\Theta
(x)=0$]. This effective control policy is equivalent to the protocol
of instant maximization through a noisy channel provided many measurement and
control actions are performed in the characteristic time of the system
evolution, which is the case we consider here.
\par

Our aim is to study the dependence of the maximum power output with the
information. On one hand, the average information transmitted through the
noisy channel is quantified in terms of the mutual information \cite{cov91}
that the controller gets from the state of the system.
Our case ---the noisy measurement of the sign of the net force--- is
equivalent to a noisy channel called the binary symmetric channel in
information theory. For this case the mutual information can be computed
(see sec. 8.1.4 of \cite{cov91}), and it is given (in bits) by
\begin{equation} \label{I}
I = H(q) - H(p),
\end{equation}
with $ H(x) = -x \log_2 x - (1-x) \log_2(1-x) $,
$q = (1-p)b + p(1-b) $ the probability that the controller receives a negative
sign, and $b$ the probability that the actual sign of the net force is
negative. Therefore, the information
$I$ that the controller gets about the system is greatly determined by the
noise level $p$ of the channel; the maximum information is
reached for $p=0$ and it is at most $1$ bit, while for $p=1/2$ the channel
becomes
completely random and no information of the system is received by the
controller. When the probability $b$ does not depend on $p$, the relation
between the noise level $p$ and the
information $I$ can be easily expanded around $ p=1/2 $ and reads
\begin{equation}\label{series}
I(p)=\frac{1}{\ln 2}\sum_{k \text{ even}}\frac{2^k}{k(k-1)}
\left[1-(1-2b)^k \right] \left(p-\frac{1}{2}\right)^k.
\end{equation}
Inverting this relation to leading order we get for $p<1/2$ the
result~\cite{cao07}
\begin{equation} \label{pI}
p \simeq \frac12 - \sqrt{\frac{I \ln 2}{8 b (1-b)}}.
\end{equation}
On the other hand, a positive power output is obtained when there is a net
flux against the load $\Fext$ that tilts the potential. In the stationary
regime the center-of-mass moves with a mean velocity $\vmedia$ and
then the average power output (work obtained per unit time) is given by
\begin{equation}\label{Pout_eq}
\Pout = \Fext \vmedia.
\end{equation}
\par

We first analyze the dependence of the power output with the information for
the case of one particle and later for the few-particle ratchet.

\section{One particle}

\begin{figure}
\begin{center}
\includegraphics [scale=0.49] {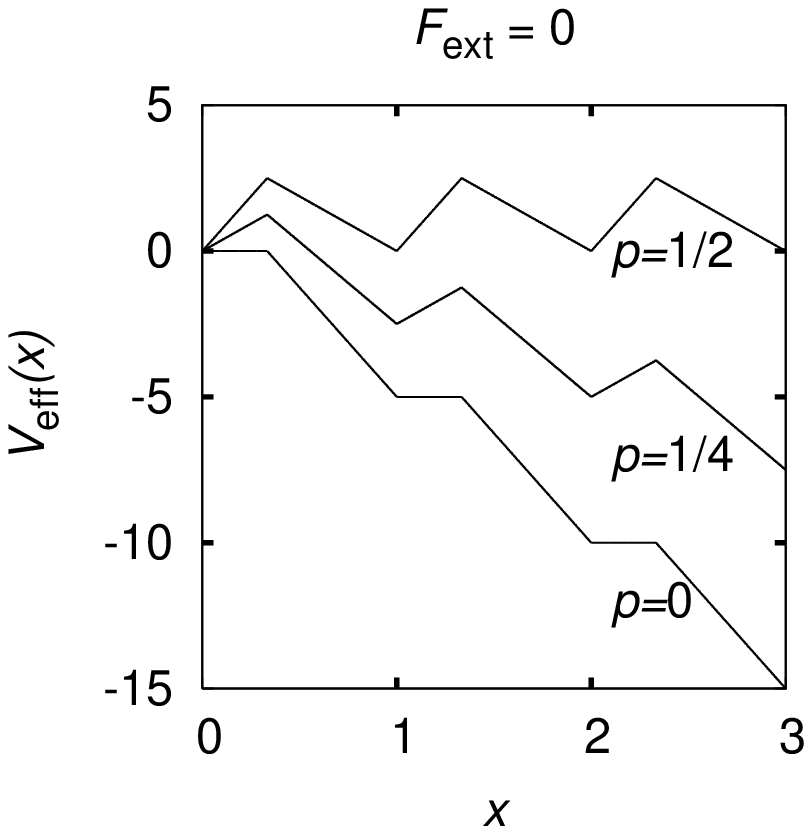}
\includegraphics [scale=0.49] {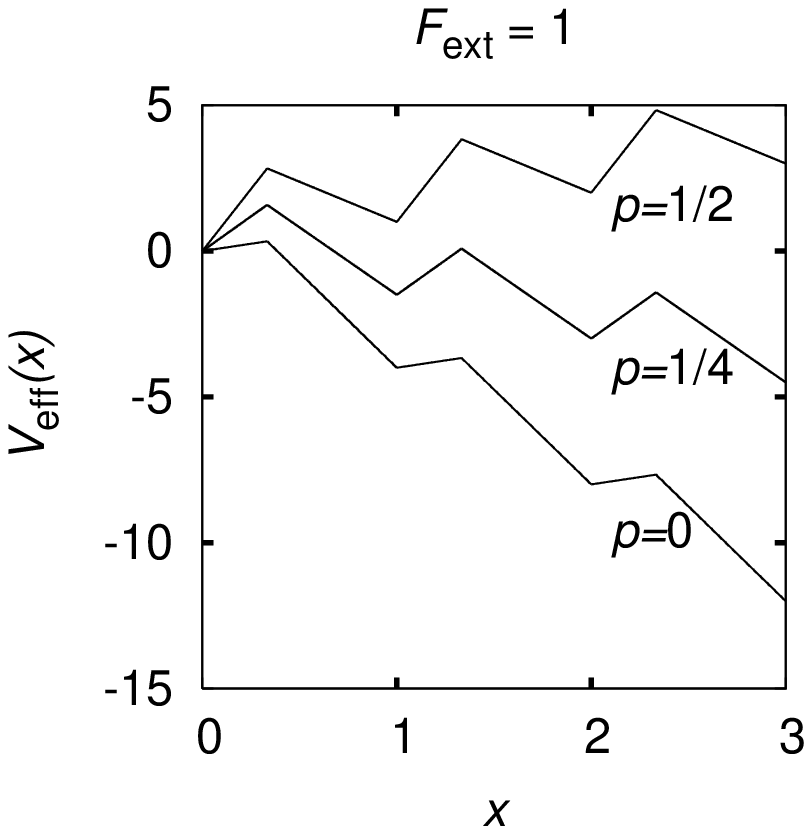}
\end{center}
\caption{Effective potentials for one particle and noise levels
$ p = 0$, $1/4$, and $1/2$ with $ V_0 = 5 $ and $ a = 1/3 $. Left panel
without external load, and right panel with external load $ \Fext
= 1 $. Units: $L=1$, $D=1$, and $k_BT=1$. Note that for $p=1/2$ the effective
potential is equal to $ V(x)/2 + \Fext x $.}
  \label{fig:Veff}
\end{figure}

We start with the one particle case ($N=1$) where an effective potential
that includes the effects of the load, the control protocol, and the binary
symmetric channel can be constructed. The system dynamics can be viewed as the
result of the action of the periodic effective force $ F_{\text{eff}}(x) =
\alpha_{\text{eff}}(x) F(x) - \Fext$ that derives from an effective potential.
Using units $L=1$ and $k_BT=1$, this effective potential can be written in terms
of $K\mathrel{\mathop :}= pV_0+\Fext a$ and
$M\mathrel{\mathop :}= (1-p)V_0-\Fext(1-a)$ as
\begin{equation} \label{Veff}
V_{\text{eff}}(x)=
\begin{cases}
\frac{xK}{a}& \mbox{if $0\leq x\leq a$},\\
K-M\frac{x-a}{1-a}& \mbox{if $a< x\leq  1$}
\end{cases}
\end{equation}
in the interval $[0,1]$, and outside
$V_{\text{eff}}(x)=V_{\text{eff}}(y)+(x-y)V_{\text{eff}}(1) $, with
$y\equiv x\mod 1$, $y\in[0,1]$.
eq.~\eqref{Veff} and fig.~\ref{fig:Veff} show that the effect of increasing
the noise level $p$ is to diminish in the effective potential the average
tilt that induces a positive flux, while the effect of increasing the load $
\Fext $ is to tilt the effective potential opposing the positive flux.
\par

Solving the stationary Fokker-Planck equation for this effective
potential the stationary mean velocity for one particle is obtained (in units
$L=1$ and $k_BT=1$):
\begin{equation}\label{vmedia}
\vone =\frac{DK^2M^2 A}{ AE-B^+ B^-},
\end{equation}
with
\begin{equation}
\begin{split}
A & \mathrel{\mathop :}= 1-e^{K-M}, \\
B^{\pm} & \mathrel{\mathop :}= [aM+(1-a)K]e^{\pm K}\\
&\quad -(1-a)K e^{\pm(K-M)} -aM,\\
E & \mathrel{\mathop :}=  a^2M^2(1-K-e^{-K})\\
& \quad +a(1-a)KM (1-e^M)(1-e^{-K})\\
&\quad +(1-a)^2K^2(1+M-e^M).
\end{split} \nonumber
\end{equation}

For $p<1/2$, there is a positive current for forces smaller than the
``stopping force'' $F_{\text{stop}}$ (the one that leads to the cancellation
of the velocity), so a work is done against the load for $\Fext\in
(0,F_{\text{stop}})$.
For $p = 1/2$, the stopping force is zero, because no positive
flux is obtained even in the absence of the external load. Our
noisy control acts instantaneously, \emph{i.e.} in a time scale
much faster than the characteristic times of the system
[$(aL)^2/(2D)$ for the diffusion time and $\gamma(1-a)^2 L^2/V_0$
for the characteristic time of the drift induced by the
potential]. Thus, for $p=1/2$ the potential $V(x)$ is randomly
switched on and off very fast and the particle just feels the
average potential. This implies that the effective potential in
absence of the load, namely $V(x)/2$ (see fig.~\ref{fig:Veff}), is
not tilted, giving a zero flux for the $ p = 1/2 $ case for zero
load. Therefore, in order to get work the noise level of the
channel should be $p\in [0,1/2)$ and  the value of
$F_{\text{stop}}$ is obtained equating eq.~\eqref{vmedia} to zero,
\begin{equation}\label{Fstop}
  F_{\text{stop}}=\frac{V_0}{L}(1-2p).
\end{equation}

\begin{figure}
\begin{center}
\includegraphics [height=5cm] {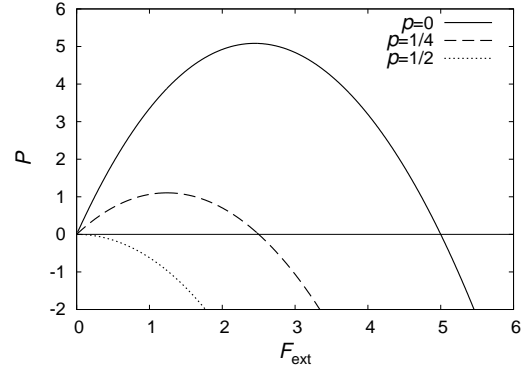}
\end{center}
\caption{Power output versus the load for $V_0=5$ and $a=1/3$ in the one
  particle case [eqs.~\eqref{Pout_eq} and~\eqref{vmedia}]. Units: $L=1$,
  $D=1$, $k_BT=1$.}
  \label{Pout}
\end{figure}

\begin{figure}
  \begin{center}
    \includegraphics [height=5cm] {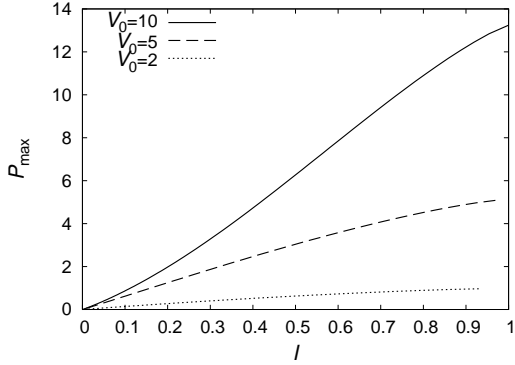}
  \end{center}
  \caption{Maximum power output in the
    one particle case as a function of the information used.  The curves are
    for heights of the potential
    $V_0=2,5,10$ and  asymmetry parameter $a=1/3$. Units: $L=1$,
    $D=1$, $k_BT=1$.}
  \label{Pmax_vs_I_varios}
\end{figure}

Substituting eq.~\eqref{vmedia} in eq.~\eqref{Pout_eq} we get the
analytical expression for the power output in the one-particle ratchet. The
dependence with the load is plotted in fig.~\ref{Pout} for noise levels $p=0$,
$p=1/4$, and $p=1/2$. The positive regions correspond to the system doing work
against the external force.
The load $\Fext^*$ that maximizes the power output lies
between 0 and $\Fstop$ and it is given by
the condition
\begin{equation}\label{max}
  \frac{\partial \Pout}{\partial \Fext}(\Fext^*)=0.
\end{equation}
In general, it is a function of the noise level of
the binary symmetric channel and it also depends on the
physical parameters of the potential, $V_0$ and $a$. The condition~\eqref{max}
gives a transcendental equation for $\Fext^*$ that
can be numerically solved in order to obtain the
maximum power output,
\begin{equation} \label{Pmax}
\Pmax=\Pout(\Fext^*).
\end{equation}
This equation gives the dependence of the maximum power with the noise level
$p$, which is related with the information $I$ through eq.~\eqref{I}. This last
equation requires to compute $b$, which can be obtained integrating over the
space interval $[0,aL]$ the stationary distribution of the Fokker-Planck
equation for the effective potential \eqref{Veff},
\begin{equation}\label{b}
\begin{split}
  b & =  \vone\left(\frac{a}{K}\right)^2\Bigg\{
    \left(  1-e^{-K}\right)
\Bigg[ 1+\\
& \quad
\frac{1-e^{-K}+(1-a)(1-e^{-M})K/(aM)}{e^{-K}-e^{-M}}
\Bigg]
-K
\Bigg\}
\end{split}
\end{equation}
(units $L=1$, $D=1$, $k_B T = 1$).
Therefore, the combination of eqs.~\eqref{I}, and (\ref{max}-\ref{b}) permits
to obtain the (implicit) exact dependence of the maximum power developed by
the Brownian motor as a function of the information gathered by the controller
(see fig.~\ref{Pmax_vs_I_varios}).
\par

We analyze now the regime of small
potentials in the one particle case. For small
potentials ($ V_0 \lesssim k_BT $) the value of the external force that
maximizes the power is also small [remember eq.~\eqref{Fstop} and the fact
that $\Fext^*<\Fstop$].  In this regime, the velocity~\eqref{vmedia} reduces
to
$\vone\simeq D(M-K)$, or, recovering units,
\begin{equation}\label{vmedia-1}
\vone\simeq(1-2p)\frac{V_0}{\gamma L}-\frac{\Fext}{\gamma}.
\end{equation}
We see that there are two contributions to the velocity: the current effect
due to the white thermal noise and the control through the binary channel,
${(1-2p)V_0}/(\gamma L)$, and the net drift due to the load,
${-\Fext}/{\gamma}$. We highlight that for small potentials (and loads) these
two effects appear uncoupled, and the result is independent of the
asymmetry of the potential. This independency of the asymmetry for small
potentials can be understood realizing that in this case the effective
potential is well approximated by a flat potential with the same average
slope, \emph{i.e.}, $ V_{\text{eff}}(x) \simeq [-V_0 (1-2p) / L + \Fext] x $.

Applying eq.~\eqref{max} to the power output computed using
eq.~\eqref{vmedia-1} we obtain
\begin{equation}\label{F*}
  \Fext^*=\frac{V_0}{2L}(1-2p)=\frac{\Fstop}{2},
\end{equation}
and then the power output is
\begin{equation}\label{p2}
  \Pmax=\frac{\Fext^{*2}}{\gamma}=\frac{V_0^2}{4\gamma L^2}(1-2p)^2.
\end{equation}
On the other hand, for small potential heights $b\simeq a$, and using
eq.~\eqref{pI} we get
\begin{equation}\label{CI}
  \Pmax\simeq R_1 I,
\end{equation}
with $R_1$ a constant that depends on the physical parameters of the system,
\begin{equation}\label{C1}
  R_1=\frac{V_0^2\ln 2}{8\gamma L^2 a(1-a)}.
\end{equation}
Notice that the dependence on the asymmetry $a$ does appear here because it
determines the relation between $p$ and $I$ [eq.~\eqref{pI}], as $ b \simeq a
$ for small potentials.

Therefore, eq.~\eqref{p2} indicates that for small potential heights and
small values of the information (\emph{i.e.}, $p \sim 1/2$) the maximum power
is approximately directly proportional to the information gathered.
In addition, we have numerically checked that eq.~\eqref{CI} gives an upper
bound of the maximum power for any potential height $V_0$ and for
any value of the information $I$.
\par

\begin{figure}
  \begin{center}
    \includegraphics [height=5cm] {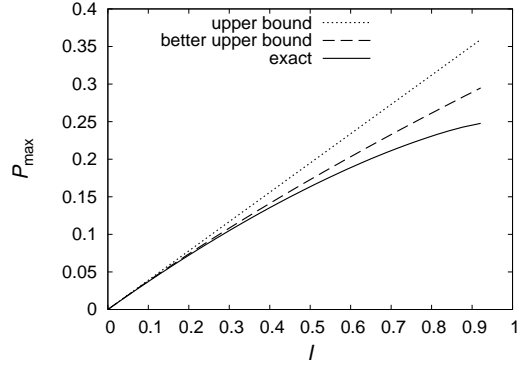}
  \end{center}
  \caption{Maximum power output in the one particle case as a function of the
    information for $V_0=1$ and $a=1/3$, and comparison with the upper
    bound~\eqref{CI} and the better upper bound~\eqref{better}. Units:
    $L=1$, $D=1$, $k_BT=1$.}
  \label{cota_I_1}
\end{figure}

A better approximation for the dependence of the maximum power output with the
information can be found using the result of inverting
eq.~\eqref{series} up to fourth order,
\begin{equation}\label{better}
\Pmax\simeq S_1 (-1+\sqrt{1+S_2 I}),
\end{equation}
with
\begin{equation}
  S_1=\frac{3V_0^2}{4\gamma L^2}\frac{1-(1-2a)^2}{1-(1-2a)^4};\;
  S_2=\frac{4}{3}\frac{1-(1-2a)^4}{[1-(1-2a)^2]^2}\ln 2,
\end{equation}
which is also an upper bound of $\Pmax$ for any potential and
information values. In fig.~\ref{cota_I_1} these upper
bounds [eqs.~\eqref{CI} and~\eqref{better}] are compared with the exact result
for $V_0=k_BT$.

\section{Few particles}
Let us now study a collective ratchet composed of a few
particles and show that the results are similar to those in the one particle
case. Summing and averaging the Langevin equations~\eqref{langevin}, the
average velocity of the center-of-mass in the stationary state can be written
as
\begin{equation}
\gamma \vmedia=\langle \alpha_{\text{eff}}(f)f\rangle -\Fext.
\end{equation}
An approximate solution can be found assuming, as in \cite{cao07,cao04}, that:
(i) the position of the particles are statistically independent, and (ii)
the probability of finding a particle in a negative force interval (for
example $[0,aL]$) is $a$. These assumptions are verified for small potentials
and small loads even in the presence of noise, and they imply that
the probability distribution for $f$ is approximately Gaussian,
\begin{equation}
\rho(f) \simeq \frac{1}{\sqrt{2\pi\Sigma^2}} \;
e^{-f^2/(2\Sigma^2)} ,
\end{equation}
with $\Sigma$ the amplitude of the fluctuations of the net force,
given by $\Sigma=\frac{V_0}{L\sqrt{a(1-a)N}}$ (see ref.~\cite{cao04}).
Following~\cite{cao07}, we get for the center-of-mass velocity
\begin{equation}\label{flux_small_approx}
  \vmedia\simeq\frac{\Sigma}{\gamma\sqrt{2\pi}}(1-2p)-\frac{\Fext}{\gamma}.
\end{equation}
This result is the sum of the center-of-mass velocity without the external
load~\cite{cao07} plus the drift $-\Fext/\gamma$ due to the external load. We
see that, like in the one particle case, these two effects
are decoupled for small potential heights. The
expression~\eqref{flux_small_approx}
agrees with the results of numerical simulations of the
stochastic evolution equations~\eqref{langevin}.
\par

Applying eqs.~\eqref{Pout_eq} and \eqref{max}, it can be shown that the
maximum power is reached for
\begin{equation}
  \Fext^*=\frac{V_0}{L\sqrt{8\pi a(1-a)N}}(1-2p)
\end{equation}
and takes the value
\begin{equation}\label{p2Cfew}
  \Pmax=\frac{V_0^2}{\gamma L^2 8\pi a(1-a)N}(1-2p)^2,
\end{equation}
or simply $\Pmax=\Fext^{*2}/\gamma$. Therefore, we also have in the regime of
few particles (and small
potentials) that the maximum power is at first approximation directly
proportional to the information,
\begin{equation}\label{Pout-few}
  \Pmax\simeq R_N I.
\end{equation}
In the previous expression,  the constant $R_N$ depends only on the physical
parameters of the system, in particular the number of particles $N$,
\begin{equation}\label{CN}
  R_N=\frac{V_0^2\ln 2}{\gamma L^2 16\pi a(1-a)b(1-b)N},
\end{equation}
where $b$ can be calculated for small potentials and loads,
\begin{equation}\label{b_few}
  b=\sum_{n>aN}^N\binom{N}{n}a^n (1-a)^{N-n},
\end{equation}
using the same assumptions that lead to the Gaussian approximation for
$\rho(f)$.
We have checked numerically that eq.~\eqref{Pout-few} is an upper bound for
the maximum power output in the few particles case.
Again, as in the one-particle ratchet, a linear upper bound has been found
for the maximum power output that the system can get using a certain amount
of information $I$.

\section{Correlation}

The previous expression of the different physical magnitudes of
the system in terms of the noise level $p$ can be recast in terms
of a correlation $C$ that we introduce in this section. The main
underlying idea is that the presence of noise in the control
induces a decorrelation between the relevant magnitudes of the
control policy. In the instant maximization of the center-of-mass
protocol~\cite{cao04} the switching of the potential only depends
on the sign of the net force, namely $\sgn f$, and the presence of
noise in the control implies that the controller does not use the
actual value $\sgn f$ but a value $\sgn \tilde f$. The correlation
between these quantities,
\begin{equation}
   {C}= \langle \sgn f \;\sgn \tilde f \rangle,
\end{equation}
can be written as
\begin{equation}
  {C} = P_{++}+P_{--}-P_{+-}-P_{-+},
\end{equation}
where $P_{+-}$ is the join probability of having $\sgn f=+1$ while
the controller receives $\sgn \tilde f=-1$, and analogously for
the other joint probabilities.  As in our system $\sgn \tilde f$
is different from $\sgn f$ with probability $p$ (the noise level)
these joint probabilities can be easily computed by noting that
$P_{-+}= bp$, $P_{--}=b(1-p)$, $P_{+-}=(1-b)p$ and
$P_{++}=(1-b)(1-p)$, with $b$ the probability of $\sgn f$ being
$-1$. Therefore, the correlation can be parameterized in terms
of the noise level $p$ as
\begin{equation}\label{cor1}
   {C} = 1-2p.
\end{equation}
In other words, the effect of the noise is to decrease the
correlation $ {C}$, which has its maximum value ($ {C}=1$) for
zero noise and its minimum ($ {C}=0$) for a completely noisy
policy, $p=1/2$.
\par

Finally, using eq.~\eqref{cor1}, the relations derived in previous
sections can be restated in terms of the correlation. For example,
eqs.~\eqref{p2} and~\eqref{p2Cfew} reads
\begin{equation}
  \Pmax=\frac{V_0^2 {C}^2}{4\gamma L^2},
\end{equation}
(one particle case), and
\begin{equation}
  \Pmax=\frac{V_0^2 {C}^2}{\gamma L^2 8\pi a(1-a)N},
\end{equation}
(few particles case).

This reformulation helps to understand the physical meaning of the
relations derived in the previous sections giving a complementary
view. In addition, it indicates that the noisy control considered
can give an effective description of other feedback ratchets with
an imperfect operation of the feedback control. For instance, this
effective description has been shown to be valid in time-delayed
feedback ratchets consisting of few particles~\cite{fei07b}, where
$f = f(t)$, $\tilde f= f (t-\tau)$ (with $ \tau $ being the
time-delay), and the correlation can be computed just from the
time series of the net force $f(t)$.

\section{Comparison with open-loop protocols}
The instantaneous maximization of the center-of-mass velocity is the optimal
protocol to maximize the power output in the one particle case for a noiseless
channel ($p=0$). Thus, we also expect this protocol to give a power
output close to the maximum possible value in the few particles case
and in the presence of noise with a memoryless protocol (note that protocols
with memory can perform error corrections).
Therefore, we expect eq.~\eqref{CI} and eq.~\eqref{Pout-few} to be upper
bounds of the maximum power output that can be obtained with a memoryless
closed-loop protocol that uses an amount of information $I$ about the system.
\par

In addition, the maximum power output obtained with open-loop protocols is
much smaller than that obtained with efficient closed-loop protocols. For
instance, for the saw-tooth potential with parameters $V_0=5k_BT$ and $a=1/3$,
the periodic protocol with optimum periods ${\cal
  T}_{\text{on}}\simeq 0.06L^2/D$ and ${\cal T}_{\text{off}}\simeq 0.04L^2/D$
gives a small maximum power $\Pmax^{\text{open}}\simeq 0.04V_0^2/(\gamma
L^2)$, which is
reached for a load $\Fext^*\simeq 0.25V_0/L$. In contrast, the closed-loop
one-particle ratchet yields a maximum power $\Pmax^{\text{closed}}\simeq
5.1 V_0^2/(\gamma L^2)$ for
$\Fext^*\simeq 2.4 V_0/L$ when it works without noise in the channel.
Therefore, the linear
eqs.~\eqref{CI} and~\eqref{Pout-few} are also good estimates of the
maximum improvement that can be attained changing from the optimal open-loop
control to a closed-loop protocol, \emph{i.e.,}
\begin{equation}
\Pmax^{\text{closed}} - \Pmax^{\text{open}} \leq R I,
\end{equation}
where $R$ is a constant depending on the system's characteristics; see
eqs.~\eqref{C1} and~\eqref{CN}.

\section{Concluding remarks}
In this article we have analyzed the relation between the
information about the state of the system used by the controller
and the power output in a feedback controlled ratchet. We have
obtained exact analytic results for one-particle ratchets, and
also approximate simple expressions for the maximum power output
in both one-particle and few-particle ratchets. Moreover, we have
found that the increase of the maximum power output when we change
from the optimal open-loop protocol to a closed-loop protocol has
an upper bound proportional to the information used by the
controller. Also an upper bound proportional to the information
was found in~\cite{tou00} for the entropy reduction in a general
closed-loop controlled system. The result obtained in the present
paper for the maximum power output is the analog upper bound of
the one found in \cite{cao07} for the flux, but with the important
difference that the upper bound for the flux was proportional to
the square root of the information.

\section*{Acknowledgments}

This work has been financially supported by grants
BFM2003-02547/FISI, FIS2005-24376-E and FIS2006-05895 from MEC
(Spain), and by the ESF Programme STOCHDYN. M.F. acknowledges
support from UCM (Spain) through grant ``Beca Complutense''.


\begin{thebibliography}{0}
\bibitem{rei02} P. Reimann,  {Phys. Rep.} {\bf 361}, 57 (2002)
\bibitem{lin02} H. Linke, {Appl. Phys. A} {\bf 75}, 167 (2002);
 P.  H\"{a}nggi, F. Marchesoni and F. Nori, {Ann. Phys.} {\bf 14}, 51 (2005)
\bibitem{bie07} M. Bier, Biosystems {\bf 88}, 301 (2007)
\bibitem{ser07}
  V. Serreli, C.-F. Lee, E. R. Ray and D. Leigh, Nature (London) {\bf 445},
  523 (2007);
  E. A. Kay, D. A. Leigh and F. Zerbetto, Angew. Chem. Int. Ed. 46, 72 (2007)
\bibitem{par02} J. M. R. Parrondo and B. J. Cisneros, {Appl. Phys. A} {\bf 75},
  179 (2002);
  H. Linke, M. T. Downton and M. J. Zuckermann, \emph{Chaos} {\bf 15}, 026111
  (2005)
\bibitem{ast94} R. D. Astumian and M. Bier, {Phys. Rev. Lett.} \textbf{72},
  1766 (1994)
\bibitem{mag93} M. O. Magnasco, {Phys. Rev. Lett.} \textbf{71}, 1477 (1993);
  H. Kamegawa, T. Hondou  and F. Takagi, {Phys. Rev. Lett.} {\bf 80}, 5251
  (1998); 
  L. Machura, M. Kostur, P. Talkner, J. \L uczka, F. Marchesoni
  and P. H\"anggi, {Phys. Rev. E} {\bf 70}, 061105 (2004)
\bibitem{ajd93} A. Ajdari and J. Prost, {C. R. Acad. Sci. Paris II}
  \textbf{315}, 1635 (1993);
  D. Suzuki and T. Munakata, \emph{Phys. Rev. E} {\bf
    68}, 021906 (2003)
\bibitem{cao04} F. J. Cao, L. Dinis and J. M. R. Parrondo, {Phys.
  Rev. Lett.} {\bf 93}, 040603 (2004)
\bibitem{din05} L. Dinis, J. M. R. Parrondo and F. J. Cao,
  {Europhys. Lett.} {\bf 71}, 536 (2005);
  M. Feito and F. J. Cao, {Phys. Rev. E} {\bf 74}, 041109 (2006)
\bibitem{cao07} F. J. Cao, M. Feito  and H. Touchette, \emph{Information and
    flux in a feedback controlled Brownian ratchet}, arXiv:cond-mat/0703492
    preprint (2007)
\bibitem{sek97} K. Sekimoto, {J. Phys. Soc. Jpn.} {\bf 66}, 1234 (1997);
 J. M. R Parrondo, J. M. Blanco, F. J. Cao  and R. Brito, {Europhys. Lett.}
 {\bf 43}, 248 (1998); 
  A. Parmeggiani, F. J\"{u}licher, A. Ajdari and J. Prost,
  {Phys. Rev. E} {\bf 60}, 2127 (1999)
\bibitem{cov91} T. M. Cover and J. A. Thomas, \emph{Elements of Information
    Theory} (John Wiley, New York, 1991)
\bibitem{fei07b} M. Feito and F. J. Cao, \emph{Time-Delayed Feedback control
    of a flashing ratchet}, arXiv:0706.1496 preprint (2007)
\bibitem{tou00} H. Touchette and S. Lloyd, {Phys. Rev. Lett.} {\bf
    84}, 1156 (2000);
  H. Touchette and S. Lloyd, {Physica A} {\bf 331}, 140 (2004)
\end{thebibliography}
\end{document}